\documentstyle[aas2pp4]{article}

\begin{document}
\title{Binary Induced Neutron-Star Compression, Heating, and Collapse}

\author{G. J. Mathews}

\affil{
University of Notre Dame,
Department of Physics,
Notre Dame, IN 46556}

\author{J. R. Wilson}
\affil{
University of Notre Dame,
Department of Physics,
Notre Dame, IN 46556 \\
and\\
University of California,
Lawrence Livermore National Laboratory,
Livermore, CA  94550}

\date{\today}
\begin{abstract}
We analyze several aspects of the recently noted neutron star collapse
instability in close binary systems.  We
utilize (3+1) dimensional and spherical numerical general relativistic
hydrodynamics to study the origin, evolution, and parametric sensitivity
of this instability.  We derive the modified
conditions of hydrostatic equilibrium for the stars
in the curved space of quasi-static orbits.  We examine the 
sensitivity of the instability to the neutron star mass and
equation of state.  We also estimate
limits to the possible interior heating and associated neutrino luminosity
which could be generated as the stars gradually compress
prior to collapse.  We show that the radiative
loss in neutrinos from this heating could exceed the
power radiated in gravity waves for several hours prior
to collapse. The possibility that the radiation neutrinos could
produce gamma-ray (or other electromagnetic) burst phenomena
is also discussed.
\end{abstract}


\section{Introduction}
\label{sec:level1}
In recent numerical studies of the relativistic hydrodynamics
of close neutron star binaries  in three spatial dimensions
(Wilson \& Mathews 1995; 
Wilson, Mathews \& Marronetti 1996, henceforth WMM96), 
it was noted that as the
stars approach coalescence they appear to experience a collapse
instability.  For an appropriate equation of state,
binary neutron stars might generally become black holes
many seconds prior to merger.  If correct, this effect could
have a significant impact on the anticipated 
gravity wave signal from neutron
star binaries near coalescence.  Such premerger collapse
might also be associated with heating, neutrino production,
and electromagnetic bursts as the released gravitational energy
from collapse is converted into thermal energy of the stars.  

Moreover, the numerical evidence that such an instability exists 
poses a number of new questions
such as the sensitivity of the instability to the specific equation
of state employed, or the intrinsic spin and masses of the stars.
One would also like to understand the time history of the
collapse and any associated electromagnetic or neutrino
emission.  

In this paper we present some new three dimensional (3D)
calculations which begin to examine these issues.
Unfortunately, however, such relativistic hydrodynamic
calculations in three spatial dimensions are computationally expensive.
A complete systematic study of this instability in three spatial 
dimensions will be long in coming.  In this paper, however,
we show that in large part this effect can
be replicated in terms of modified one-dimensional spherical
relativistic hydrodynamics.   We show that the relativistic effects 
of placing  stars in a close binary can be 
approximated by adding a term involving an average 
Lorentz-like factor which increases the effective
gravitational forces on the stars.  The collapse
observed in the three dimensional calculations can
 be understood in this
one-dimensional framework and one can survey easily the
sensitivity of this effect to parameters characterizing the
binary and the neutron star equation of state.  

We can also follow the possible precollapse compression and 
 heating of the neutron star material.  This provides 
a framework in which to model the possible associated
neutrino and electromagnetic signals such as gamma-ray bursts.
We show that significant heating and neutrino emission
is possible as the stars gradually compress before
they reach the collapse instability.
During the heating epoch
the associated neutrino and electromagnetic
radiative losses may dominate over the power loss
from gravitational radiation.

\section{Field Equations}
Our method of solving the field equations in three spatial
dimensions  was discussed in Wilson \& Mathews (1995) and  WMM96.
 Here we present a brief review of
some features relevant to the present discussion.

We start with the slicing of spacetime into a one-parameter
family of hypersurfaces
separated by differential displacements in
time-like coordinates as defined in the (3+1)
formalism (Arnowitt, Deser \& Misner 1962; York 1979).

Utilizing Cartesian $x, y, z$ isotropic  coordinates,
proper distance is expressed
\begin{equation}
ds^2 = -(\alpha^2 - \beta_i\beta^i) dt^2 + 2 \beta_i dx^i dt + \phi^4
\delta_{ij}dx^i dx^j
\end{equation}
where the lapse function $\alpha$
describes the differential lapse of proper time between two hypersurfaces.
The quantity  $\beta_i$ is the shift vector denoting the shift in space-like
coordinates between hypersurfaces.  For an orbiting binary,
$\beta_i$ is dominated by the orbital motion of the system plus
a small contribution from frame drag (WMM96).
In the frame of one star in the one dimensional
calculations described here, most of the effect of the $\beta_i$
is transformed away.  

The curvature of the 3-geometry is described by a position
dependent conformal factor $\phi^4$ times a flat-space Kronecker delta.
We refer to this gauge choice 
as the {\it Conformally Flat Condition} or CFC.  For a static
system, the vanishing of the Weyl tensor in three dimensions
guarantees (cf. Weinberg 1972) that there exists a conformally flat solution
to the Einstein equations.  One must be careful, however,
not to overimpose symmetry conditions (e.g. Cook et al. 1996).
For a dynamic system, one can always impose conformal flatness
as an initial condition.  There are, however, nonzero time derivatives
of the spatial metric and extrinsic curvature which can begin to
introduce off-diagonal elements of $\gamma_{ij}$ as the system evolves.  
In particular, the imposition
of conformal flatness excludes gravity wave 
information contained in the transverse traceless components of the metric.
However, as discussed below, several recent studies have
indicated that this approach is still an excellent approximation
when a comparison with exact results can be made.

The implementation of the CFC means that
we solve the
constraint equations of general relativity  at each time
as though there were a fixed distribution of matter.
We then evolve the hydrodynamic equations to the next time step
against this background metric.
Thus, at each time slice we can obtain
a solution to the relativistic field equations and
information on the hydrodynamic evolution. Information on the generation of
gravitational radiation can then be obtained from a multipole expansion
(Thorne 1980) of the transverse traceless components of the metric.

It is important to appreciate that at each time slice
a numerically valid solution to the field equations is obtained.
In this way the strong field properties of the system
are included.  For this reason,  this approach  is a significant
improvement over a post Newtonian approach (which is also conformally flat at low 
order: cf. Appendix A).
The hydrodynamic variables respond to these fields.
An approximation we have made herein is the
neglect of an explicit coupling of the gravity waves. These, however,
contribute negligibly to the metric, stress energy tensor,
or hydrodynamic evolution
(WMM96). When desired, this coupling can be
added via a multipole expansion.

We reduce the solution of the equations for the field variables
$\phi$, and $\alpha$ to simple Poisson-like equations in flat space.
We begin with the Hamiltonian constraint equation (York 1979)
which reduces to (Evans 1985; WMM96),
\begin{equation}
\nabla^2{\phi} = -4\pi\rho_1.
\label{phi}
\end{equation}
In the Newtonian limit the source term is dominated  by the
proper matter density $\rho$.  In the strong field of the
orbiting binary, however, $\rho$ must be enhanced by 
a generalized curved-space Lorentz factor $W$ [cf. Eq.~(11)].  
This derives directly from the occurrence of the four velocity in
the stress energy tensor.
There are also contributions from the internal energy 
density $\epsilon \rho$, pressure $P$, and extrinsic curvature $K_{ij}$.  
Thus we write, 
\begin{eqnarray} \rho_1&=&{\phi^5 \over 2}\biggl[\rho W^2 +
\rho \epsilon \biggl( \Gamma W^2 - \Gamma+1\biggr) \nonumber \\
&&+ {1 \over 16\pi} K_{ij}K^{ij}\biggr]~,
\label{rho1}
\end{eqnarray}
where $\Gamma$ is
an adiabatic index from the equation of state as  defined below.
Similarly, the lapse function is determined from,
\begin{equation}
\nabla^2(\alpha\phi) = 4\pi\rho_2~~,
\label{alpha}
\end{equation}
\begin{eqnarray}
\rho_2 &= &{\alpha \phi^5 \over 2}\biggl[\rho (3W^2-2)+
\rho \epsilon[ 3\Gamma (W^2+1)-5]\nonumber \\
&& + {7  \over 16\pi} K_{ij}K^{ij}\biggr]~.
\label{rho2}
\end{eqnarray}

In WMM96 it was shown that the collapse instability
can at least in part be traced to the effect of the Lorentz-like  factor
$W$ on the source density for $\phi$ and ($\alpha \phi$).
In WMM96 and below it is shown that terms which scale as $(W^2 - 1)$
also enter into the hydrodynamic equations in a way which enhances
the gravitational force on each star.
In Appendix A we suggest
that, even in a post-Newtonian approximation, such
terms might cause the effective gravitational
potential to be deeper than it would be for static isolated stars.

Regarding the reliability of the {\it CFC} as an approach to this 
problem we note that a recent study (Cook, Shapiro \& Teukolski 1996) 
has shown than an axially symmetric CFC  
approximation is quite good when computed 
physical observables are compared with the
exact results for axisymmetric extremely rapidly 
rotating neutron stars.  This is the simplest
system for which an exact metric begins to differ
from a {\it CFC} metric.  

In another  recent application to the nonaxisymmetric case of
orbiting binaries Reith \& Sch\"afer (1996) have shown that
an expansion using this metric is identical to
a post-Newtonian expansion for terms of order $(v/c)^2$.
The first deviation appears in terms of
order $(v/c)^4$.  However, we find that the
deviations are small.  The expressions in their paper
are in terms of a dimensionless parameter $\nu \equiv m_1 m_2 /(m_1 + m_2)^2$.
It is common in post-Newtonian expansions to compare with
a Schwarzschild orbit for which $\nu = 0$.  However,
for neutron star binaries $\nu = 0.25$ is most appropriate.
For equal-mass neutron-star binaries $\nu = 1/4$ exactly. 
Even for  $m_1/m_2 = 2$ (a relatively large asymmetry for neutron stars)
$\nu = 0.22$.  For $\nu =  0.25$ 
the difference  between the conformally flat and post Newtonian
$(v/c)^4$ correction for the perihelion advance 
is about is 4.5\%.  The $(v/c)^4$ term
for the momentum differs by about 2.8\%, and the angular momentum
term differs by 24.1\%.  Since $(v/c)^4 \sim 10^{-4}$
for the binaries considered here, these differences in
the two-body dynamics
are probably insignificant.  Note also, that in the present application
we compute an exact instantaneous numerical solution to the Einstein equations
using this metric and do not rely upon an expansion which may deviate
in individual terms but still provide the correct results.  Note also,
that the principle effect we are investigating here  (that due to the
$W^2 - 1$ terms) is of order $(v/c)^2$ (see Appendix A) for which the
post-Newtonian and conformally flat terms agree exactly.  
Also, note that the effect described here
is a relativistic effect which completely dominates [see below] over
the possible stablizing influence of Newtonian tidal distortion 
as proposed in Lai (1996).  
The effect we describe was  not considered in that paper.

\section{Relativistic Hydrodynamics}
\label{hydro}

To solve for the fluid motions of the system in curved spacetime
it is convenient to use an Eulerian fluid description (Wilson 1979).
We begin with the perfect fluid stress-energy tensor, which in covariant
form can be written,
\begin{equation}
T_{\mu\nu} = (\rho + \rho \epsilon + P)U_\mu U_\nu + P g_{\mu \nu}~~,
\end{equation}
where $\epsilon$ is the internal energy per gram, $P$ is the pressure, and
$U_\nu$ is the four velocity.

By introducing a set of Lorentz contracted state variables it is possible to
write the relativistic hydrodynamic equations in a form which is
reminiscent of their Newtonian counterparts.  
The hydrodynamic state variables are:
The coordinate baryon mass density, 
\begin{equation}
D = W \rho~~,
\end{equation}
the coordinate internal energy density, 
\begin{equation}
E = W \rho \epsilon~~,
\end{equation}
the spatial three velocity,
\begin{equation}
V^i = \alpha {U_i \over \phi^4 W} - \beta^i~~,
\label{three-vel}
\end{equation}
and the coordinate momentum density,
\begin{equation}
S_i = (D + \Gamma E) U_i~~.
\label{momeq}
\end{equation}
The  Lorentz-like factor $W$ is 
\begin{equation}
W = \alpha U^t~ = \biggl[ 1 + {\sum_{i = 1}^3{U_i^2} \over \phi^4}\biggr]^{1/2},
\label{weq}
\end{equation}
and the EOS index $\Gamma$ is 
\begin{equation}
\Gamma = 1 + {P \over \rho \epsilon}~.
\end{equation}
Note that in flat space ($\alpha = \phi  = 1$),  $W$ reduces to the
usual special-relativistic Lorentz factor.

In terms of these state variables, the hydrodynamic equations are as follows:
The equation for the conservation of baryon number $(\rho U^\mu)_{;\nu} = 0$ takes
the form,
\begin{equation}
{\partial D\over\partial t}  =  -6D{\partial \log\phi\over\partial t}
-{1\over\phi^6}{\partial\over\partial x^j}(\phi^6DV^j)~~.\
\end{equation}
The internal energy equation is derived from
$T_{\mu~;\mu}^{~\nu} = 0$,
\begin{eqnarray}
{\partial E\over\partial t}  =&&  -6\Gamma E{\partial \log\phi\over\partial t}
-{1\over\phi^6}{\partial\over\partial x^j}(\phi^6EV^j)\nonumber \\
&& -  P\biggl[{\partial W\over \partial t} +
{1\over\phi^6}{\partial\over\partial x^j}(\phi^6 W V^j)\biggr]~~.
\end{eqnarray}
The spatial components of the momentum conservation 
condition ($T_{i~;\mu}^{~\nu} = 0$) takes the form,
\begin{eqnarray}
{\partial S_i\over\partial t}& = & -6 S_i{\partial \log\phi\over\partial t}
-{1\over\phi^6}{\partial\over\partial x^j}(\phi^6S_iV^j)
-\alpha{\partial P\over \partial x^i} \nonumber \\
& + & 2\alpha(D+\Gamma E)(W - {1\over W}){\partial \log\phi\over\partial
x^i} + S_j {\partial \beta^j \over \partial x^i} \nonumber \\
& - & W(D + \Gamma E){\partial \alpha \over \partial x^i} ~,
\label{hydromom}
\end{eqnarray}
where for the present stability analysis we have set the
radiation reaction term (WMM96) to zero.

Regarding the stability of our treatment of numerical relativistic 
hydrodynamics, we
have applied a number of standard tests (e.g.~shock tubes,
pressureless collapse, etc.) as noted in
WMM96.  One important test in the present context is
that of stable stars in a stable
orbit (with no radiation reaction).
We have found that for such systems, equilibrium configurations
are obtained after a small fraction of an orbit.  When the 
velocity damping is removed there is no discernible change of
the stars for several orbit periods.  This illustrates an advantage of
the shifted grid which we employ.  There is essentially no matter
motion with respect to the grid once a stable equilibrium has
been achieved.  Hence, numerical stability can be maintained 
for a long time.  

Another important test is that of a single 
nonrotating star on the three-dimensional spatial grid.
We find that  the equilibrium gravitational mass as a function of
central density agrees with the spherical hydrostatic 
Tolman-Oppenheimer-Volkoff
equilibrium gravitational mass as a function of
central density to within a fraction of a percent.
We also find that a dynamical instability ensues once the
stellar mass exceeds the maximum hydrostatic mass as expected.  

In isotropic coordinates, the condition of hydrostatic equilibrium for
the stars ($dS^i/dt = 0$, $V^i = 0$, 
$\partial \log{\phi}/\partial t = 0$) can be inferred 
from  equation (\ref{hydromom}),
\begin{eqnarray}
{\partial P \over  \partial x^i}&&= -(\rho + \rho \epsilon \Gamma)
\biggl( {\partial \log{\alpha} \over \partial x^i} \ 
- {U_j \over \alpha}{\partial \beta^j \over \partial x^i} \nonumber \\
&& + \biggr[ {\partial \log{\alpha} \over \partial x^i} -
  2 {\partial \log{\phi} \over \partial x^i}\biggr] (W^2 - 1) \biggr)~~.
\label{hydrostat}
\end{eqnarray}
Some discussion of the relative magnitude of the
terms in Eq.~\ref{hydrostat} is useful.
The first term with ${\partial \log{\alpha}/\partial x^i}$ is
the relativistic analog of the  Newtonian gravitational force.
In the Newtonian limit $\alpha \rightarrow 1 - GM/r$. Hence
$-{\partial \log{\alpha}/\partial x^i} \rightarrow GM/r^2$. 
 In Eq. (\ref{hydrostat}) there are two ways in which the
effective gravitational force increases as $W$ exceeds unity.
One is that the matter contribution to the source density 
for $\alpha$ or $\phi$ 
is increased by factors of $\sim W^2$ [cf. Eqs. (\ref{rho1},\ref{rho2})].
The more dominant effect is that from the terms in Eq.~\ref{hydrostat}
which scale as $(W^2 - 1)$.  These terms result from the affine connection 
terms
$\Gamma^\mu_{\mu \lambda} T^{\mu \lambda}$ in the covariant differentiation
of $ T^{\mu \nu}$.  These terms have no Newtonian analog but 
describe a general relativistic
increase in the curvature gravitational force as the specific kinetic energy
of the system increases.
 This increase in effective gravity as the stars  approach each other
can be thought of as a correction to the
Newtonian gravity which scales as  
$(W^2-1)$ times the Newtonian gravity.  This $(W^2 - 1)$
factor can be thought of as
a kind of  specific kinetic energy [cf.~Eq. (\ref{weq})] from
the orbital  motion of the binary. The extra 
${\partial log{\phi}/\partial x^i}$ term  further increases the
effect by a factor of 2.  This factor comes from $\phi^2 \sim (1/\alpha)$.
A further increase of binding arises from the $K^{ij}K_{ij}$ terms in the
field sources, but these terms are much smaller than the
$W^2-1$ contributions.

In our shifted spatial grid, the fluid three velocities $V^i$ are
nearly zero.  Hence we can use Eq.~(\ref{three-vel}) to
find $U_i$ as a function of $\beta^i$.   We can also replace $\beta^i$ 
in Eq.~(\ref{three-vel}) with the dominant
contribution from orbital motion, 
 $\bar\beta \approx \bar R \times \bar \omega$,
where $R$ is the coordinate distance of the stars from the center of mass.
Hence, Eqs.~(\ref{three-vel}) and (\ref{weq})
give,
\begin{eqnarray}
\langle W^2 \rangle &\approx &
 {1 \over (1 -  \omega^2 R^2 \phi^4/\alpha^2) }\nonumber \\
 &\approx &{1 \over (1 - v^2 /c^2 ) }~~,
\end{eqnarray}
where $c= \alpha/\phi^2 < 1$ is the coordinate light speed.

In our  simulations, an effective
velocity of $(\omega R/c) \approx 0.25$ is obtained
for the last stable orbit of 1.45 M$_\odot$ stars.
In the 3D calculations the average 
$\langle W^2 - 1 \rangle$ typically rises up to $\sim$ 5\% before 
the orbit becomes dynamically unstable.  Thus, we estimate that
before orbit instability,
the effective hydrostatic gravitational force on the
stars is increased by $\sim 10$\% over that of stationary non-orbiting
stars for which $\langle W^2 - 1 \rangle =0$.  This 
increased gravitational force increases the central densities
as the stars approach and can induce collapse.

It is  of interest to compare the magnitude of the $(W^2-1)$ correction to
the Newtonian gravity with the magnitude of the Newtonian tidal
energy which would tend to stabilize the star.  In Lai (1996) it was
estimated that the Newtonian tidal energy should scale as
\begin{equation}
\Delta E_{tidal} \approx -\lambda {G M^2 R^5 \over r^6}
\end{equation}
where for neutron stars $\lambda \approx 0.1$, $M$ is the mass
of a star, $R$ is the neutron star radius, and $r$  is 
the orbital separation.  In contrast, the correction
to the Newtonian self gravity from the motion of the 
stars in a binary is 
\begin{equation}
\Delta E_{GR} \approx 2(W^2 - 1){G M^2 \over R}
\end{equation} 
Taking the ratio of these two contributions,
we find,
\begin{equation}
{\Delta E_{tidal} \over \Delta E_{GR}}  \approx {\lambda \over 2(W^2 - 1)}
 \biggl({R \over r}\biggr)^6 \sim 10^{-4}
\end{equation}
where we have used typical values near collapse (cf. Table 1)
of $R/r \sim 0.2$ and $(W^2 -1) \sim 0.05$.  Thus, 
the effect of the increased relativistic gravitational force is expected
to dominate over the stabilizing tidal distortion by 
about 4 orders of magnitude.   In a similar analysis, we estimate
that even for white dwarfs near merger, this relativistic 
increase in the gravitational
energy exceeds the stabilizing Newtonian tidal energy.

The centrifugal term in Eq.~(16) is dominated
by the contribution from orbital motion 
$U_j (\partial \beta^j/\partial x^i) \approx U_j \omega \sim 10^{-8}$.
This term varies little over an individual star and
inside a star this term is small compared to the
${\partial \log{\alpha}/\partial x^i}$ term.  Hence,
this term can be neglected in the discussions
of stellar stability.  It is important, however,
for determining the orbits and gravity wave frequency (WMM96).

\section{Equivalent Spherical Model}
To better understand the relativistic effects described herein,    
it is useful to reduce the
hydrodynamic equations to an approximate spherical model.
Such a model can also 
be used to make a schematic survey of the sensitivity of collapse
to EOS parameters and possible interior heating. 
From Eq.~(\ref{hydrostat}) we see that the configuration of
each star can be described by a modified version
of the familiar equation of hydrostatic equilibrium.
This is true as long as the contribution of orbital motion to $(W^2-1)$
can be treated as a constant factor and the $K_{ij}K^{ij}$ and centrifugal
terms can be ignored. 
Of course, $W^2$ is not constant over the stars. However, 
in our three dimensional calculations it is observed
to vary little over the volume of a star.  

Contours of constant $(W^2-1)$  from a three-dimensional calculation
are shown in Fig. \ref{fig1}.  From this we deduce that it
is not a bad approximation to replace $(W^2 - 1)$ in the 
source equations and hydrodynamical equations with
\begin{equation}
(W^2 - 1) \approx (W_r^2 - 1 + \langle W_0^2 - 1 \rangle)~~,
\end{equation}
where $W_r$ is the contribution from radial motion inside a star, 
and $\langle W_0^2 - 1 \rangle$ is an approximately constant factor which
accounts for the influence of orbital motion in the curved space-time
 of the binary.
The equilibrium and stability of a binary star can then be approximated
using a one-dimensional description.

\placefigure{fig1}

However, since the metric variables $\alpha$ and $\phi$ depend
upon the density distribution, it is not possible to directly
compute the hydrostatic equilibrium of the star.  Instead,
the star must be evolved hydrodynamically (with damping)
as $\langle W_0^2-1 \rangle$ is increased to obtain the new equilibrium
configuration.  Hence, we construct a modified spherical
hydrodynamic model as follows:

For a given distribution of mass and energy, the
Poisson equations (\ref{phi}) and (\ref{alpha}) for $\phi$ and $\alpha$
can be integrated directly.  The only difference is that
the source terms now become
\begin{equation}
\rho_1 \approx {\phi^5 \over 2}\biggr(\rho(1 + \epsilon
 + \epsilon \Gamma (W_r^2 - 1 + \langle W_0^2 - 1\rangle)\biggr)~~,
\end{equation}
and
\begin{eqnarray}
\rho_2 & \approx & {\alpha \phi^5 \over 2}\biggr(3\rho(1 + 
 \epsilon \Gamma) (W_r^2 - 1 + \langle W_0^2 - 1 \rangle)\nonumber \\ 
&& + \rho(1 + \epsilon + 6 \epsilon(\Gamma - 1))\biggr)~~.
\end{eqnarray}

The hydrodynamic equations become:
\begin{equation}
{\partial D\over\partial t}  =  -6D{\partial \log\phi\over\partial t}
-{1\over\phi^6 r^2}{\partial\over\partial r}(\phi^6r^2DV^r)~~.\
\end{equation}
The equation for internal energy conservation becomes,
\begin{eqnarray}
{\partial E\over\partial t}  =&&  -6\Gamma E{\partial \log\phi\over\partial t}
-{1\over\phi^6 r^2}{\partial\over\partial r}(\phi^6 r^2  EV^r)\nonumber \\
&& -  P\biggl[{\partial W_r\over \partial t} +
{1\over\phi^6 r^2}{\partial\over\partial r}(\phi^6 W_r  r^2 V^r)\biggr]~~.
\end{eqnarray}
The momentum equation is
\begin{eqnarray}
{\partial S_r\over\partial t}& = & -6 S_r{\partial \log\phi\over\partial t}
-{1\over\phi^6 r^2}{\partial\over\partial r}(\phi^6 r^2 S_rV^r)
-\alpha{\partial P\over \partial r} \nonumber \\
& + & 2\alpha(D+\Gamma E)\biggl({(W_r^2 - 1 + \langle W_0^2-1 \rangle)\over W_r}\biggr)
{\partial \log\phi\over\partial
r}  \nonumber \\
& - & (W_r^2  + \langle W_0^2-1 \rangle)(D + \Gamma E){\partial \alpha \over \partial r} ~~, 
\label{hydromom1}
\end{eqnarray}
where we have neglected the centrifugal term as noted above.

The calculations reported here were performed
on an Eulerian grid in which
the star is resolved into about 100 radial  zones.

\section{Equation of State}
\label{EOS}

A key part of the calculations presented here is the
use a realistic neutron star equation of
state (EOS).  
The orbital calculations presented in WMM96 used
the zero temperature, zero neutrino chemical potential 
EOS   from the 
supernova numerical model of  Wilson \& Mayle (1993), Mayle, Tavani 
\& Wilson (1993).  
Calculations made with this
EOS for a model of supernova 1987A give an explosion energy of
$1.5 \times 10^{51}$ ergs, consistent with observation.
Also, the neutrino spectra and time of neutrino emission are in good
agreement with the IMB (Bionata et al. 1987)
and Kamiokande neutrino detections (Hirata et al. 1987).
These models also reproduce the desired abundance
distribution of $r$-process heavy elements 
in the baryon wind from the proto-neutron star (Woosley et al. 1994).
The maximum mass neutron star for this EOS as converted
to a zero temperature version for our present studies
is  M$_C =$ 1.70 M$_\odot$.
An important point is that with an EOS which would allow
a higher mass neutron star,  Wilson  \& Mayle (1993) were not able to
obtain satisfactory results.
 
In WMM96 only cold equilibrium configurations
were computed.  However, in the present
work we wish to examine the possible heating of the stars
as they collapse.  Hence,
we include finite temperature effects in the EOS.
The electron fraction is small for neutron stars $Y_e << 1$.  
Hence, for the heating calculations  of interest here,
we can relate the temperature to  the internal energy by assuming         
a nonrelativistic fermi gas of neutrons.  

We wish to analyze the sensitivity of the collapse instability to
the neutron star equation of state.
To do this we diminish the maximum mass
achievable for a given equation of state by
imposing a maximum value for the index $\Gamma$
at high density.  From this maximum, the 
adiabatic index asymptotes to 2 at high density to guarantee causality.

We find a maximum neutron star mass of  M$_C =$ 1.55, 1.64, and 1.70
for $\Gamma_{max} =$  2.297, 2.346, 2.470, respectively.
This range of masses is consistent with (and even slightly above)
the upper range of the observed upper mass limit for
neutron stars.   Finn (1994) 
has assigned a lower limit of 1.15 to 1.35 M$_\odot$ 
and an upper limit of
1.44 to 1.50  M$_\odot$ at the $1 \sigma$ (68\%) confidence level.
At the $2 \sigma$ (95\%) confidence level the upper limit increases to
1.43 to 1.64  M$_\odot$. In an independent approach,
Bethe and Brown (1995)  have recently argued from nucleosynthesis
constraints that the maximum neutron star mass is 1.56 M$\odot$.
They also point out that if kaon condensation is taken into account 
the critical mass may only be 1.50 M$_\odot$.
If the maximum observed stellar mass were as low as  
the $1 \sigma$ upper limit,
i.e.  1.50  M$_\odot$, it could be
that almost all neutron star binaries would collapse before coalescence.

With the present state of knowledge of the nuclear equation
of state at high density, however, it is still possible that the
maximum neutron star mass could be significantly greater than
1.70 M$_\odot$.  That is, the observed low mass limits
may only be an artifact of the way in which neutron stars
are formed  in type II
supernovae rather than a limit from the EOS.
 We have made some preliminary studies
 of stars with  M$_{G} = 1.45$ M$_\odot$ for an EOS 
with M$_C =$ 1.85 M$_\odot$.
We have not observed collapse 
before the orbit instability is reached.  It seems likely that 
for a sufficiently stiff EOS that merger will occur  as two
neutron stars as considered in many Newtonian and
post Newtonian simulations (e.g. Rasio \& Shapiro 1992; 
Zhuge, Centrella, McMillan 1994; Janka \& Ruffert 1996).

\section{Summary of 3D results}
\subsection{Summary of Previous Results}
 In WMM96 orbit calculations were made for two
$1.70$ M$_\odot$ baryonic mass neutron stars for
an EOS for which the gravitational mass
in isolation was  M$_{G} = 1.45$
M$_\odot$ and the critical neutron star mass was  M$_C      = 1.70$ M$_\odot$.
Orbit solutions were sought for three separate
values of angular momentum.   The neutron stars were taken
to be corotating initially, although it was noted that
relativistic effects subsequently induce some fluid motion 
in the stars relative to the corotating frame.  In the present
study we have not considered the more realistic
possibilities of initial neutron-star
spins or non equal masses.  Such systems contain less symmetry
and require a larger computational effort
 which we leave for future work.

The first calculation was made with an orbital angular momentum
of $2.2 \times 10^{11}$ cm$^2$.  The stars settled down into
what appeared at first as a stable orbit, but later 
(less than one complete orbit) the stars began to slowly
spiral in.  
The next calculation was made with an angular momentum of $2.3 \times 10^{11}$
cm$^2$ for which the orbit appeared stable.
However, after about 1 to 2 revolutions the central
densities were noticed to be rising.  By the end of the
calculation the central baryonic densities had continuously
risen to about $2.7 \times
10^{15}$ g cm$^{-3}$ ($\approx 10$ times nuclear matter density)
which is near the maximum density for a stable
neutron star for an EOS with M$_C$ = 1.70.  
It appears that neutron stars of this mass
range and the adopted equation of state may continue to collapse
as long as the released gravitational energy can be dissipated.
For this orbit the stars are at a separation distance
of $d_p / m = 9.5$, far from merging.  
By the time the calculation was ended, the
minimum $\alpha$
had diminished to 0.379 and $\phi^2$ risen to 2.05
corresponding to a minimum coordinate light speed of 0.18.  

  A third   calculation was made with the angular momentum increased
  to $2.7 \times 10^{11}$ cm$^2$.  As can be seen in Table \ref{paramtable}
and  the stars at this separation
$d_p / m = 12.4$ seemed both stable and in a stable orbit. However,
with only a slight increase in baryonic  mass (M$_b$ $1.598 \rightarrow 1.620$ 
a collapse ensues.

\placetable{paramtable}
\placetable{paramtabl2}
\placetable{paramtabl3}
\placetable{paramtabl4}

\subsection{New 3D Results}
In the present work, these results are supplemented with additional
3D calculations for initially corotating stars.
The new results are obtained with better  resolution
and an improved treatment of boundary conditions.  
These calculations have been
run several times longer than in WMM96 so that  for 
cases where the instability occurs we have
followed the collapse to higher densities and stronger fields.

These new results  along with the previous results
are summarized in the Tables \ref{paramtable} through \ref{paramtable4}.
For the star with M$_G \approx 1.45$ M$_\odot$ in isolation,
we have added calculations at intermediate orbital angular
momenta of $2.5 \times 10^{11}$
and $2.6 \times 10^{11}$ cm$^2$ here we have run for
much longer times and further into the collapse.  We find that even at 
$2.6 \times 10^{11}$ cm$^2$ the stars collapse while still at a distance
of over 50 km apart and with a gravity wave frequency of only 250 Hz.
The collapse instability appears to onset between
 $2.6$ and  $2.7 \times 10^{11}$ cm$^2$ as $\langle W^2 - 1 \rangle 
\rightarrow \sim 0.05$  (depending upon the baryon mass).  

The rate of energy and
angular momentum loss generally increases as the compression proceeds and 
$\alpha$ becomes small.  We note, however, that for unstable stars or
orbits, the systems are no longer in quasi-equilibrium orbits.  
Since the radiated energy and momentum are sensitive functions of
separation and $\omega$, the computed values of energy and angular 
momentum loss become unreliable as an average estimate. This
is the reason for the lack of monotonicity in the $\dot E$
values of Tables \ref{paramtable} - \ref{paramtable4}.
Eventually the system
approaches two black holes and is no longer
well describable in our framework.
From the rates of energy and angular
momentum loss for these calculations we make a crude 
estimate (WMM96) that
a delay of about 5 seconds occurs between the collapse instability and 
the orbit instability for stars with this EOS and $M_G = 1.45$  M$_\odot$.
This would have interesting consequences on the gravity wave 
or electromagnetic signals as discussed below.

We have also run 3-dimensional
calculations  for stars which would have $M_G = 1.40$  M$_\odot$
and 1.35  M$_\odot$ in isolation (Tables \ref{paramtable3} 
and \ref{paramtable4}).
For these systems the EOS was varied as well as the angular momentum.
Results from these calculations are summarized in Tables 2 - 4.
We found as expected that the 1.40  M$_\odot$ stars are stable
for lower angular momenta than the 1.45  M$_\odot$ stars.
However, stars of this mass will still collapse for an EOS which is more soft.
For example, the 1.40  M$_\odot$ star will collapse for 
$J = 2.6 \times 10^{11}$ cm$^2$
if the upper mass limit from the EOS is reduced to 1.55  M$_\odot$.  
We have found that with the M$_C = 1.70$  M$_\odot$  EOS it is necessary
to reduce the mass to $M_G = 1.35$  M$_\odot$ before the stars can survive to
final orbit plunge without collapsing first (see Table 3).

\subsection{Connection to 1D Calculation}

  One of the concerns in the 3D calculations of WMM96 was whether
the stars and field variables were sufficiently resolved to produce reliable 
numerical results.  In the 3D calculations, the spatial resolution
only provided $\sim 10-15$ zones in radius across a star.  
In the 1D calculations,
however, one can easily provide many radial zones.  
We have made a survey of
various physical quantities such as
the central extrema in $\alpha$, $\phi$, and  $\rho$,  as well as
the total gravitational mass M$_G$ as a function of the 
number of radial zones.   
We have found that there is no significant difference in the 
field variables, hydrodynamics variables,
 or gravitational mass as the radial zoning is increased
from 15 to 200 zones.  Even for only 10 radial zones
the error in mass only rises to $\sim$1\%.  Hence, the zoning in the
3D calculations discussed here and in WMM96 is probably not
a significant source of uncertainty.

We wish to explore the reliability of the 1D model as the Lorentz-like 
factor $\langle W_0^2 - 1 \rangle $ is increased. Figure \ref{fig2} shows the proper 
central baryon density as a function of $\langle W_0^2 - 1 \rangle$.  Results are 
given for two different EOS's; one for which M$_C      = 1.64$ M$_\odot$;
and one with M$_C = 1.70$ M$_\odot$.  Also shown for
comparison are central densities as a function of
average values  for $(W^2 - 1)$ from 3D calculations.
We see that the basic trend of increasing central density with
increasing orbital motion is reproduced, although the 
central density in the 1D calculations is
about 2\% higher than the 3D calculations for the
same average 
$\langle W_0^2 - 1 \rangle$ factor.  This suggests that replacing the distribution
in $W$ with a mass weighted average value slightly overestimates
the effect.  Nevertheless, this approach is sufficiently accurate 
to apply to the schematic parameter study of interest here.

\placefigure{fig2}

\section{Heating}

  In WMM96
the  released binding energy 
was assumed to be deposited only in increased fermi energy.
Any thermal excitation  was assumed to be radiated away so that the stars
remained cold.
However, it is not necessarily true that the input energy
above the increasing fermi energy
goes into thermal energy  or that it is efficiently radiated away.  
If this energy were not
dissipated, the stars could simply oscillate about
the equilibrium rather than collapse.  
We argue, however, that it seems most likely that
such oscillations would be
quickly damped relative to the time scale for
inspiraling.  Initially, the radial changes will
be quite small, and the coupling of radial motion to
thermal excitation
could occur, for example, via star quakes in analogy with
observed pulsar glitches.  As the
rate of energy release becomes more rapid
and the crust melts,
we speculate that the coupling of radial modes
with the orbital motion, nonradial fluid motion, and tidal forces will lead to
a complex excitation of higher modes and shocks
which could further heat the star and increase the entropy.
Also, the coupling of radial modes with the magnetic fields
could damp the oscillations.
Eventually, as the stars become hot enough,
$T \sim 1$ MeV, neutrino viscosity will serve
to damp the radial motion, but this would be late in the evolution. 

As these dissipative processes come into play it seems
plausible that significant thermal energy could be excited.
If the thermal energy is efficiently radiated away, then
the stars will remain near zero temperature
and the previous calculations are valid.  However,
it is also possible that the energy may not be
radiated away as rapidly as it is released.
In which case the 
damping will be converted into both increased fermi energy and
thermal energy.  An upper limit to the temperature of the
star would be that corresponding to no
radiation during the compression.   In the present
work we estimate the 
possible heating and radiation of the stars
as they adjust to the changing orbit $\langle W_0^2 - 1\rangle$ factor
and tidal forces.

If there is sufficient heating, the 
stars may produce an associated neutrino and/or
electromagnetic signal as they compress.
Hence, it is of interest to estimate the possible heating
of the stars as released binding energy is converted to
internal energy.

One can make a simple estimate (Mathews \& Wilson 1996) 
for the heating from the change in gravitational binding
energy as the stars compress in the 3D calculations.  
From Table \ref{paramtable},
as $J$ changes by $4 \times 10^{10}$  cm$^2$
in the three-dimensional numerical calculations
the angular momentum loss rate is
$\dot J \sim 1$ cm.  The
time to radiate this change in
$J$ is $\Delta J/J \sim 1.33$ sec.
From the change of binding energy with
central density for single stars for a particular EOS,  
it is possible (Mathews \& Wilson 1996) 
to estimate the energy available for heating of the stars
after increasing the fermi energy.
This has an associated change in
baryonic mass, hence it is only an order of magnitude estimate
which we improve upon below.

Nevertheless, this change in binding energy could correspond to
a release of as much as
$6 \times 10^{52}$ ergs in thermal energy and a heating rate
of $5 \times 10^{52}$ ergs s$^{-1}$ per star.  The corresponding
average energy over the stars could be $2 \times 10^{19}$ ergs g$^{-1}$.  
If this energy were injected into a star on the verge
of collapse, (central density of $2.42 \times 10^{15}$ g cm$^{-3}$)
it would heat the core to
a temperature $T \approx 45$ MeV (assuming that the core
has the heat capacity of a degenerate Fermi gas of neutrons.)

This much heating could lead to copious neutrino emission
and may provide a framework in which to produce
gamma-ray or X-ray bursts.
Hence, there is motivation to numerically study
this possible heating.  We do this in the spherical calculations
by imposing an energy conserving damping  term
in the equations of motion.
This  damping relaxes
the stars to their new equilibrium.  
The damped kinetic energy is added as
internal energy.  By integrating the
rate at which kinetic energy is damped into
Fermi and thermal  energy, and estimating the fraction which
can be subsequently radiated away,
we get a measure of the
possible heating of the star before collapse.

\section{Results}

\subsection{Analysis of the Collapse Instability}
  With the above 1D approximation
to the effects of orbital motion, we can make
 a systematic study of stellar stability
as a function of mass, EOS, and $\langle W_0^2 - 1\rangle$ factor.
First we set
$\langle W_0^2 - 1\rangle = 0$ and run a hydrodynamic calculation 
at zero temperature with velocity damping until
an equilibrium configuration is achieved
for a given gravitational mass and EOS.  Then we increase
$\langle W_0^2 - 1\rangle$ in small increments and evolve the star
hydrodynamically with conservative damping. That is, the
damped kinetic energy is added to the internal and thermal energy as 
the calculation proceeds.  

For stable stars,
we generally observe that the kinetic energy at first rises
to a maximum and then damps to zero in the hydrodynamic simulations.  
For a star which
has reached the collapse instability, however, the kinetic
energy first rises to a maximum then falls to a minimum
and then begins to rise again as the collapse ensues.
Hence, we define  the collapse instability for this systematic
study as the point at which the
radial kinetic energy begins to increase with time rather than
relaxing  to zero.  We wish to analyze the heating and neutrino
emission up to this point. Once the instability is reached, the
stars quickly collapse and much of the subsequent heating or neutrino emission
becomes lost below the event horizon.

Figures \ref{fig3}-\ref{fig5} show the central value of the
lapse $\alpha$ and the released
gravitational energy (in units of 10$^{53}$ ergs) $E_{53}$ as 
$\langle W_0^2 - 1\rangle$ increases from zero to the point at which 
dynamical collapse is evident.  

Instability in these calculations is also
manifest by $\alpha$ decreasing rapidly once
it falls below some critical value. In these calculations this instability
occurs as $\alpha \rightarrow 0.4-0.5$ depending upon
the value of $(W^2 - 1)$, the initial mass, and the EOS.  
A similar phenomenon has been noted
in previous studies of spherical stars in the
isotropic gauge.  It was noted (Wilson 1979)
that once $\alpha < 0.5$ unstable collapse generally ensues,
but of course those calculations had no $(W^2 - 1)$ effect.
We see the same  $\alpha < 0.5$ 
limit as the mass is increased with $(W^2 - 1) = 0$.  The
central $\alpha$ in the 3D orbit calculations near collapse is 
also about 0.4.

The size of the Lorentz factor at instability and the amount of input thermal 
energy increases with initial gravitational mass M$_G$ and critical mass
M$_C$ of the EOS as one would expect.
The released energy rises  approximately
quadratically with $\langle W_0^2 - 1\rangle$.  The total
energy released up to the instability point increases with decreasing
mass and increasing M$_C$ of the EOS.  

\placefigure{fig3}
\placefigure{fig4}
\placefigure{fig5}

\subsection{Temperature and Neutrino Luminosity}

The absolute surface neutrino luminosity will depend upon
details of  the neutrino transport from the interior.  
It should scale, however, with surface temperature according to a
Steffan Boltzmann law, $L_s \propto T_s^4$,
and the total luminosity should scale with the
surface radius $r$ and the lapse $\alpha_s$ at the surface,
$ L_{tot} \propto  L_s  r^2 \alpha^2_s~~.$

If the luminosity
is not as great as the rate at which
thermal energy is added to the stars, then central temperatures 
and the associated neutrino luminosity could be quite high.
To estimate  the luminosity and temperature as a function of time 
consider the Newtonian angular frequency,
\begin{equation}
\omega^2 \approx { m \over r^3}~~,
\label{omega2}
\end{equation}
 and deceleration
due to quadrupole radiation (cf. Blanchet et al. 1995),
\begin{equation}
\dot \omega = {96 \over 20} m^{5/3} \omega^{11/3}~~.
\label{omegadot}
\end{equation}
The evolution of $\omega$ from earlier times
($t < 0$) up to a value $\omega_0$ can be estimated
by integrating Eq.~(\ref{omegadot}).
\begin{equation}
\omega = {\omega_0 \over (1 - (64/5) m^{5/3} \omega_0^{8/3} t)^{3/8}}~~.
\label{omegat}
\end{equation}
From Eq.~(\ref{omega2}) the time
dependence of $(m/r)$  is 
\begin{equation}
{ m \over r} = {(m \omega_0)^{3/2} \over
(1 - (64/5) m^{5/3} \omega_0^{8/3} t)^{1/4}}~~.
\end{equation}
Since $(W^2 - 1)$ can be thought of
as a kind of specific kinetic energy, it should scale
as $m/r$ in a stable Keplerian orbit.
\begin{equation} 
W^2 - 1 = {\sum U_i^2 \over \phi^4} \propto {m\over r}~~.
\end{equation}
From this we get an approximate time history
\begin{equation}
\langle W_0^2 - 1\rangle = {\bar{W}_{3D}^2 - 1 \over
(1 - (64/5) m^{5/3} \omega_0^{8/3} t)^{1/4}}~~,
\label{wnorm}
\end{equation}
where Eq.~(\ref{wnorm}) is normalized to give the average
$\langle W_{3D}^2 - 1\rangle$ factor from the 3D calculation
which has an angular velocity  
$\omega_0$.

From this relation of $\langle W_0^2 - 1\rangle$ 
as a function of time it is possible
to now construct a possible picture of the luminosity 
as a function of time.  Ideally, one would like
to model the detailed thermal neutrino production
and transport as released gravitational energy is
deposited in the interior.  Although we have begun 
such a calculation, a detailed modeling of the neutrino
transport is quite challenging and it will take
some time before a systematic study can be completed.
Nevertheless, we can gain qualitative insight
into the signal expected from a simple schematic model 
as follows:

From (Figs. \ref{fig3}-\ref{fig5})
we note that the total input thermal energy
grows quadratically with $\langle W_0^2 - 1\rangle$, 
\begin{equation}
E_{in} \propto \langle W_0^2 - 1\rangle^2~~.
\label{ethw}
\end{equation}
To convert this thermal energy into a luminosity
we assume   that the neutrino flux through any 
surface at radius $r$ should scale as,
\begin{equation}
F_\nu \propto \biggl({r^2 T_0^2 \over \rho \kappa_0 T^2}\biggr)
{ d(T^4) \over d r} \propto
{r^2 \over \rho} T \ {d T \over d r} ~~,
\label{flux1}
\end{equation}
where $\kappa_0$ is the opacity evaluated at $T_0$.

The net neutrino flux passing through $r$
is due to neutrinos diffusing throughout the volume interior to
$r$.  To approximate the effective temperature of the flux
passing through $r$ we assume that the thermal energy
is deposited uniformly in mass throughout the star.
The neutrino flux at any radius $r$ is then taken to be 
proportional to the mass interior to $r$,
\begin{equation}
F_\nu \propto m(r)~~,
\label{flux2}
\end{equation}
where,
\begin{equation}
m(r) = 4 \pi \int_0^r \rho r'^2 dr'~~.
\end{equation}
Equations (\ref{flux1}) and (\ref{flux2}) define
an effective neutrino temperature profile with radius,
(cf. Fig \ref{fig6}),
\begin{equation}
T(r)  = A \biggl[\biggl({\int_0^r m(r')dr' \over r'^2}\biggr)
 - \biggl({\int_0^R m(r')dr' \over r'^2}\biggr)\biggr]^{1/2} ~~,
\label{Tr}
\end{equation}
where $A$ is a normalization constant.
We then solve equations (\ref{flux1}) through (\ref{Tr})
under the boundary condition that $T = 0$
outside the star, and by 
equating the total thermal energy at any given time to the
integrated thermal internal energy per gram $\epsilon (T)$
for a given temperature and density
profile in our the equation of state, i.e.
\begin{equation}
\int_0^R \epsilon (T) 4 \pi r^2 \rho dr = E_{th}~~.
\label{enorm}
\end{equation}

The radial temperature profile of  a 1.40 M$_\odot$ star just as 
the collapse instability is reached, is shown
in Figure \ref{fig6}  for the equations of
state with M$_C = 1.70$.  For this EOS  central temperatures
as high as $T \approx 70$ MeV are possible.  

The neutrino flux is then calculated using Eq.~(\ref{flux1}) with
the proper coefficients included.  The flux near the surface
then gives the luminosity.
The luminosity can be used to then
define an effective neutrino luminosity temperature,
\begin{equation} 
L_{\nu} = 4 \pi R^2 {a c T_{eff}^4 \over 4} \biggl({11 \over 4}\biggr)~~.
\label{ltot}
\end{equation}

\placefigure{fig6}

As the stars compress, released gravitational energy can be
deposited as internal energy to be radiated away by neutrinos.
The rate of accumulated thermal energy
is then given by the balance between the rate at which gravitational
energy is deposited as thermal energy 
from the contraction of the stars $\dot E_{in}$
and the rate of energy loss $L_\nu$ by neutrinos. 

We find that the temperature profile as determined above
is consistent with a scaling $E_{th} \sim T^{2}$, where 
$T$ is the evaluated from (Eq.~\ref{Tr}) near the
the surface.  This scaling arises because the system 
can be approximated as a degenerate nucleon gas.
Also, from the flux scaling (Eq.~\ref{flux1})
we find that $dT/dr \propto T$,
so that $L_\nu \sim T^2$.  
Hence, both $E_{th}$ and $L_{\nu}$
scale with the surface temperature $T^2$, and we can write 
\begin{equation}
L_{\nu} \propto T^2 \approx k E_{th}~~.
\end{equation}
The evolution of the thermal energy can then be written,
\begin{equation}
\dot E_{th} = \dot E_{in} - k E_{th} ~~.
\end{equation} 
The constant $k$ is evaluated from Eq.~(\ref{flux1}) above, 
and the rate of deposited thermal
energy is evaluated from Eqs.~(\ref{wnorm}) and (\ref{ethw}).
\begin{equation}
\dot E_{in} \approx {d E_{in} \over d \langle W_0^2 - 1\rangle} 
{d \langle W_0^2 - 1\rangle \over dt}~~.
\end{equation}
The analytic solution for the heating of the star then becomes,
\begin{equation}
E_{th} \approx e^{-k t} \int \dot E_{in} e^{k t'} dt'~~.
\end{equation}

Figure \ref{fig7} shows 
the  estimated neutrino luminosity
$L_{\nu}$ (in units of $10^{53}$ ergs sec$^{-1}$), 
the total accumulated internal energy, $E_{th}$ (in units of $10^{53}$ ergs),
and the rate of gravitational energy release, $\dot E_{in}$
(in units of $10^{53}$ ergs sec$^{-1}$),
for a 1.40 M$_{\odot}$ star with the $M_C = 1.70$ EOS.  These
quantities are plotted  as a function
of time for the last 1.5 sec of a star with M$_G$ = 1.40 M$_\odot$
and an EOS with M$_C$ = 1.70 M$_\odot$ where the time scale is defined by
Eq.~\ref{omegat}.

The total luminosity can become quite significant as the 
collapse instability is approached.  
About 4 sec before collapse, the neutrino luminosity from each star rises
above  10$^{51}$ ergs sec$^{-1}$.
The luminosity exceeds  10$^{52}$ ergs sec$^{-1}$
about 0.5 sec before collapse.  The combined neutrino luminosity from
the two stars ultimately reaches nearly 10$^{53}$ erg sec$^{-1}$ 
before collapse.  This is comparable to the neutrino emission
from type II supernovae, but in this case the emission
is from bare neutron stars.  

\placefigure{fig7}

In the quadrupole approximation (cf. Thorne 1980; WMM96), 
the gravity wave luminosity $\dot E_{GW}$
scales with the square of the  $(l+1)$th time derivative of the mass
quadrupole moment,
$\dot E_{GW} \sim Q^2 \omega^6 \approx (m/r)^5$.   Thus, we write,
\begin{equation}
\dot E_{GW} = {\dot E_{GW}^{3D} \over (1 - (64/5) m^{5/3} 
\omega_0^{8/3} t)^{5/4}}~~.
\end{equation}
From this we estimate that the power (and angular momentum) lost in
neutrinos will exceed the energy loss in gravity waves for
roughly 3 hours before collapse.  This means that the late evolution
up to collapse may not be determined by the rate of 
gravitational radiation
but by the hydrodynamics and heat transport of the compressing
neutron stars.  

If the radiative momentum loss dominates at early times one may
wonder whether it could be observed as an increased slow down rate
 in the orbit of a known binary pulsar.
We have estimated how much the orbit period of the binary pulsar
PSR-1913+16 would
be affected.    Damour \& Taylor (1991) have determined that 
the ratio of the observed orbit period change to the
general relativistic prediction is 
$\dot P^{obs}/ \dot P^{GR}
= 1.0081 \pm 0.0022$(galactic) $\pm 0.0076$(observational).  
We estimate that any orbit period change from 
compressional heating is at least two orders of magnitude
below the observational error.

\section{Conclusion}

We have made a survey of the compression, heating,  and 
collapse of neutron stars in close binaries.  In particular,
we have developed a schematic model to describe
when the collapse instability may
occur as a function of initial neutron star  mass and the EOS.
 We have also analyzed the
possible heating of the neutron star interiors
as the stars approach the collapse instability.
We find that the stars may
obtain quite high thermal energy and neutrino luminosity 
in the final seconds before collapse.  This could have significant
implications both for gravity wave and neutrino astronomy
as follows:

\subsection{Implication for Gravity Wave Detectors}

The analysis here indicates that 
the radiative neutrino luminosity
could exceed the gravity luminosity for hours
prior to the collapse instability.  
If so, this could have a profound influence
on the inferred gravity wave signal.  The loss of orbital 
angular momentum  due to neutrinos
and electromagnetic radiation will be considerably 
greater than that of two cold stable neutron stars.
The merger will occur on a shorter time scale and the
gravity wave signal will be dominated by the dynamics of
heating and thermal radiation and not the gravity wave amplitude
up to the point of the instability.

Once the collapse instability is reached, we estimate that the 
formation of one or two black holes will occur rather abruptly. 
After collapse, however, the system may not appear simply as
two black holes in vacuum.  As has been observed
in supernova calculations for some time  (cf. Mayle \& Wilson 1988;
Wilson \& Mayle 1993)
this much neutrino radiation is likely to ablate  electron-positron
pairs together with baryonic
material from the surface of the stars.
Baryons ejected in this wind are
likely to be present after collapse and may interact with the
orbiting objects.  To some extent they will provide
material to accrete onto the remaining members (neutron star or black holes)
 of the binary.
They may also provide a damping medium which could accelerate
the decay of the orbit.  Thirdly, this
hot wind material may provide a medium in which to anchor
the magnetic field lines of the precollapse stars (cf. Wilson 1975;
Ruffini \& Wilson 1975; Damour et al. 1978).

We speculate that these effects may serve to accelerate the
merger of the two black holes.  The interaction of the stars
with this medium may affect the dynamics of
the black hole inspiral unless the 
material is ejected with sufficiently high velocity.
Clearly, this is an area which warrants
further investigation.  If our speculation is correct then
the gravity wave signal becomes a probe
of the EOS, hydrodynamics, and thermodynamics of the neutrons
stars as they approach and pass through this collapse instability.

\subsection{Implications for Gamma-Ray Bursts}

The possibility that gamma-ray bursts could be generated
by neutrino emission from coalescing 
neutron stars has been speculated upon for some time.
Recently,  Janka \& Ruffert (1996) have made post-Newtonian
hydrodynamics calculations of neutron star mergers and
included the neutrino emission therefrom.
They find high luminosities, but the time scales are so short ($\sim
$ msec) that they conclude that it will be difficult to model
gamma-ray bursts by neutron star mergers.  This short time
scale stems from the time scale for mergers. 
This difficulty
is avoided, however,  in our model in which the time scale is set by
the gradual compression of the stars.  We estimate
similar luminosities, but in our model the neutrino
luminosity endures for much longer times,  thus rendering
the possibility of a gamma-ray burst more plausible.

We have shown that significant 
heating and associated
neutrino luminosity is possible in the last seconds before the
collapse instability.  This poses some interesting possibilities
for cosmological models of gamma ray bursts.  The thermal emission
of neutrinos provides an environment for the
generation of an $e^+-e^-$
 pair plasma by $\nu \bar \nu$ annihilation
around the stars.  The neutrino 
 emission is occurring in the deepening gravitational
well of the two stars.  Their interactions will be
enhanced by the curved space around the neutron stars.
Furthermore, the region between the stars may provide an environment
for the build up of neutrino and matter flux and the
production of a pair plasma as desired in some gamma-ray
burst scenarios (e.g. Piran \& Shemi 1993).

In addition to the collapse-induced neutrino emission itself,
the escaping neutrinos are likely to generate 
a neutrino heated baryon wind from the stars (Mayle 
\& Wilson 1988; Wilson \& Mayle 1993). 
 Unlike in supernovae, the velocity of this wind can be be quite high, particularly
later in the evolution as the neutrino luminosity grows.
This later emission of the high velocity wind could interact
with matter emitted previously producing shock heating in
environments of relatively low optical depth far from the stars.
The interactions themselves may contribute to the production 
of a pair plasma.  

As a preliminary test of this scenario
we ran a calculation of neutron stars
instantly heated such that the  surface temperature was $\sim 5 $ MeV.
We then followed the neutrino and matter transport  using the numerical
supernova model of Mayle \& Wilson (1993).  
We observed a blow off of the outer layer
($\sim 10^{-5}$ M$_\odot$) of the neutron star.  This material
was accelerated  to a speed corresponding to a
relativistic $\gamma$-factor of $\sim10$.  One
possibility is that this high speed matter interacting
with magnetic fields and/or interstellar clouds might produce $\gamma$-rays.

Finally, we also note that after
collapse, the previously ejected material will continue to
experience heating
either by accretion onto the black holes or by 
ram pressure from the orbiting stars.  
Once present, this plasma might become anchored 
to magnetic field lines around the precollapse stars
(Ruffini \& Wilson 1975; Damour et al. 1978).
The interactions and magnetic recombination of these
field lines could also contribute to heating and pair plasma
production.

All of these processes may be occurring in the background of the
remaining orbiting binary system from times prior to
collapse until the final merger to a single black hole.
This orbit period may  lead to an underlying 
millisecond substructure in associated burst signals
possibly consistent with observations.

Clearly, this is an area which also warrants more investigation.
Work along this line is underway to explore such effects as
a possible framework in which to model cosmological 
gamma-ray bursts.

\acknowledgments

The authors wish to thank P. Marronetti for useful discussions
and contributions to this work.
Work at University of Notre Dame
supported in part by DOE Nuclear Theory grant DE-FG02-95ER40934
and by NASA CGRO Grant NAG5-3123 and NAG
Work performed in part under the auspices 
of the U.~S.~Department of Energy
by the Lawrence Livermore National Laboratory under contract
W-7405-ENG-48 and NSF grant PHY-9401636.

\begin{deluxetable}{lrrrrr}
\tablewidth{0pc}
\tablecaption{Parameters characterizing the orbit calculations 
for stars with  $M_G^0 \approx  1.45$ M$_\odot$ and
for an EOS with M$_C = 1.70$.  }
\tablehead{
\colhead{$J$ ($10^{11}~ cm^2$)}      & \colhead{$2.2$} &
\colhead{$2.3$} & \colhead{$2.5$} &
\colhead{$2.6$} & \colhead{$2.7$}}

\startdata
$M_B$ (M$_\odot$)&$1.598$&$1.598$&$1.620$&$1.620$&$1.598$\nl
$M_G$ (M$_\odot$)&$1.416$&$1.420$&$1.322$&$1.317$&$1.423$\nl
$f$ (Hz)&$410$&$310$&$280$&$250$&$267$\nl
$d_P$ (km)&$39.4$&$40.6$&$50.3$&$50.7$&$53.0$\nl
$\rho_{max}$ ($10^{15}$ g cm$^{-3}$)&$2.03 $&$2.70 $&$3.58$&$2.92$&$1.93$\nl
$\langle W_{3D}^2 -  1 \rangle$ &$-$&$-$&$0.050$&$0.052$&$0.043$\nl
$\alpha_{min}$ &$0.440$&$0.379$&$0.283$&$0.288$&$0.463$\nl
$\phi^2_{max}$ &$1.90 $&$2.05 $&$1.78$&$2.68$&$1.84 $\nl
$\dot E$ (M$_\odot$ sec$^{-1})$&$0.016$&$0.0040$& $0.00048$&$0.00047$&$0.0061$\nl
Orbit  & Unstable & Stable & Stable & Stable& Stable\nl
Stars & Unstable & Unstable & Unstable& Unstable & Stable\nl 
\label{paramtable}
\tablecomments{M$_G$
 is the total gravitational mass of the binary divided by 2. Also, $f$ is
the gravity wave frequency, i.e. twice the orbit frequency.}
\enddata
\end{deluxetable}

\begin{planotable}{lrrrrrrr}
\tablewidth{0pc}
\tablecaption{Parameters characterizing the orbit calculations at the final
edit for stars with  $M_G^0 = 1.40$  (M$_B$ = 1.548) M$_\odot$ in isolation and
for equations of state with M$_C = 1.70$,  1.64, and 1.55 M$_\odot$.}
\tablehead{
\colhead{$J$ ($10^{11}~ cm^2$)}      & \colhead{$2.5$} &
\colhead{$2.6$} & \colhead{$2.5$} &
\colhead{$2.6$} & \colhead{$2.5$} &
\colhead{$2.6$} & \colhead{$2.7$} 
 \\[.2ex]
\colhead{EOS - $M_C$}          & \colhead{1.70} &
\colhead{1.70}      & \colhead{1.64} &
\colhead{1.64}      & \colhead{1.55} &
\colhead{1.55}  & \colhead{1.55}}

\startdata
$M_G$ (M$_\odot$)&$1.300$&$1.306$&$1.299$&$1.306$&$1.285$&$1.299$&$1.310$\nl
$f$ (Hz)&$270$&$287$&$272$&$251$&$256$&$260$&$233$\nl
$d_P$ (km)&$62.1$&$64.8$&$61.7$&$65.6$&$55.3$&$60.3$&$68.2$\nl
$\rho_{max}$ ($10^{15}$ g cm$^{-3}$)&$1.82$&$1.79$ &$1.82$&$1.86$&$2.95$
&$2.36$&$1.98$\nl
$\langle W_{3D}^2 - 1\rangle$ &$0.0384$&$0.0365$&$0.0387$&$0.0354$&
$0.0434$&$0.0298$&$0.0339$\nl 
$\alpha_{min}$ &$0.435$&$0.443$&$0.426$&$0.444$&$0.266$&$0.396$&$0.440$\nl
$\phi^2_{max}$ &$2.05 $&$2.02 $&$2.08$&$2.02$&$2.91$&$2.18$&$2.02$\nl
$\dot E$ (M$_\odot$ sec$^{-1})$&$0.0016$&$0.00155$&$0.00156$&$0.00141$
&$0.000548$ &$0.00113$&$0.00112$\nl
Orbit  & Stable & Stable & Stable & Stable & Stable & Stable & Stable \nl
Stars & Stable & Stable  & Stable & Stable & Unstable & Unstable & Stable \nl
\label{paramtable2}
\enddata
\end{planotable}

\begin{planotable}{lrrrrrr}
\tablewidth{0pc}
\tablecaption{Parameters characterizing the orbit calculations at the final
edit for stars with  $M_G^0 = 1.35$  (M$_B$ = 1.49) M$_\odot$ in isolation and
for equations of state with M$_C$ = 1.70 M$_\odot$.}
\tablehead{
\colhead{$J$ ($10^{11}~ cm^2$)}      & \colhead{$1.90$} &
\colhead{$1.95$} & \colhead{$2.00$} &
\colhead{$2.05$} & \colhead{$2.10$} &
\colhead{$2.15$}  
 \\[.2ex]}

\startdata
$M_G$ (M$_\odot$)&$1.225$&$1.231$&$1.238$&$1.243$&$1.248$&$1.252$\nl
$f$ (Hz)&$413$&$388$&$365$&$346$&$330$&$312$\nl
$d_P$ (km)&$46.4$&$48.4$&$50.8$&$52.8$&$54.8$&$57.6$\nl
$\langle W_{3D}^2 - 1\rangle$ &$0.0531$&$0.0497$&$0.0459$&$0.0431$&
$0.0408$&$0.0388$\nl
$\alpha_{min}$ &$0.399$&$0.413$&$0.428$&$0.440$&$0.449$&$0.457$\nl
$\phi^2_{max}$ &$2.22 $&$2.16 $&$2.10$&$2.06$&$2.02$&$1.99$\nl
$\dot E$ (M$_\odot$ sec$^{-1})$&$0.0038$&$0.0034$&$0.0032$&$0.0030$&$0.0029$
&$0.0026$\nl
Orbit  & Unstable & Stable & Stable & Stable & Stable & Stable \nl
Stars & Unstable & Stable  & Stable & Stable & Stable & Stable \nl
\label{paramtable3}
\enddata
\end{planotable}

\begin{planotable}{lrrrrrrr}
\tablewidth{0pc}
\tablecaption{Parameters characterizing the orbit calculations at the final
edit for stars with  $M_G^0 = 1.35$  (M$_B$ = 1.49) M$_\odot$ in isolation and
for equations of state with M$_C$ = 1.55 M$_\odot$.}
\tablehead{
\colhead{$J$ ($10^{11}~ cm^2$)}      & \colhead{$2.10$} &
\colhead{$2.15$} & \colhead{$2.20$} &
\colhead{$2.30$} & \colhead{$2.40$} &
\colhead{$2.50$} & \colhead{$2.60$} 
 \\[.2ex]}

\startdata
$M_G$ (M$_\odot$)&$1.238$&$1.246$&$1.255$&$1.262$&$1.268$&$1.274$&$1.279$\nl
$f$ (Hz)&$270$&$287$&$272$&$251$&$256$&$260$&$233$\nl
$d_P$ (km)&$50.6$&$55.0$&$57.8$&$62.5$&$66.8$&$70.1$&$74.4$\nl
$\langle W_{3D}^2 - 1\rangle$ &$0.0476$&$0.0427$&$0.0379$&$0.0350$&
$0.0319$&$0.0293$&$0.0272$\nl
$\alpha_{min}$ &$0.392$&$0.424$&$0.452$&$0.465$&$0.480$&$0.493$&$0.503$\nl
$\phi^2_{max}$ &$2.23 $&$2.10 $&$2.01$&$1.96$&$1.91$&$1.86$&$1.83$\nl
$\dot E$ (M$_\odot$ sec$^{-1})$&$0.00022$&$0.0023$&$0.0022$&$0.0019$&$0.0016$
&$0.0013$&$0.0011$\nl
Orbit  & Stable & Stable & Stable & Stable & Stable & Stable & Stable \nl
Stars & Unstable & Stable  & Stable & Stable & Stable & Stable & Stable \nl
\label{paramtable4}
\enddata
\end{planotable}

\begin{figure}
\caption{Contours of $W^2 - 1$ in the $Z = 0$, $X-Y$ plane from the three-dimensional
calculations  for M$_G$ = 1.40 neutron stars with $J = 2.6 \times 10^{11}$
cm$^2$. Contour decrease from a maximum of $W^2 - 1 = 0.04$
in steps of 0.005.
 }
\label{fig1}
\end{figure}

\begin{figure}
\caption{ 
Central proper baryon density $\rho_c$ for 1D models (solid lines)
  as a function of  $\langle W_0^2 -1\rangle$ for a star 
with M$_G$ = 1.40 M$_\odot$.  Lines are drawn for
equations of state which give a maximum
neutron star mass of M$_C$ = 1.70 M$_\odot$ and 1.64 M$_\odot$ as labeled.  
For comparison, the points show numerical results from the 3D calculations 
at various values of $\langle W_0^2 -1\rangle$ 
for the M$_C$ = 1.70 M$_\odot$ EOS (squares) and the
M$_C$ = 1.70 M$_\odot$ EOS (circles).
}
\label{fig2}
\end{figure}

\begin{figure}
\caption{
Central values of the lapse function $\alpha$ and the
released gravitational energy in units of 10$^{53}$ ergs $E_{53}$
as $\langle W_0^2 -1\rangle$ increases from zero to the collapse instability
point.  These calculations use an EOS which gives a maximum
neutron star mass of M$_C$ = 1.55 M$_\odot$.  The different
curves are labeled by their associated
gravitational mass.
}
\label{fig3}
\end{figure}

\begin{figure}
\caption{
Same as Figure \ref{fig3}, but for an EOS which gives a maximum
neutron star mass of M$_C$ = 1.64 M$_\odot$.
}
\label{fig4}
\end{figure}

\begin{figure}
\caption{
Same as Figure \ref{fig3}, but for an EOS which gives a maximum
neutron star mass of M$_C$ = 1.70 M$_\odot$.
}
\label{fig5}
\end{figure}

\begin{figure}
\caption{
Radial temperature profile just as 
the instability is reached $(\langle W^2 -1 \rangle = 0.055)$
 of a 1.40 M$_\odot$ star 
for an EOS which gives a maximum
neutron star mass of M$_C$ = 1.70 M$_\odot$.
}
\label{fig6}
\end{figure}

\begin{figure}
\caption{
Estimated neutrino luminosity
$L_{\nu}$  (in units of $10^{53}$ ergs sec$^{-1}$), 
the total accumulated internal energy, $E_{th}$
 (in units of $10^{53}$ ergs),
and the rate of gravitational energy release, $\dot E_{in}$
 (in units of $10^{53}$ ergs sec$^{-1}$),
for a 1.40 M$_{\odot}$ star with the $M_C = 1.70$ EOS.  These
quantities are plotted  as a function
of time for the last 1.5 sec of a star with M$_G$ = 1.40 M$_\odot$
and an EOS with M$_C$ = 1.70 M$_\odot$.}
\label{fig7}
\end{figure}

\newpage
\appendix

\section{Post Newtonian Analysis of the ($W^2-1$) Effect}

The effect described in this paper stems from a numerical
solution to the full
Einstein field equations in the approximation that the three metric
remains conformally flat.
We have interpreted the apparent stellar collapse as the result of
a general relativistic increase  in the strength of the gravitational 
acceleration which causes the stars to compresses  and heat as the orbit shrinks.
 At least in part, this increased gravity involves
terms which scale with $(W^2 - 1)$.  Since $(W^2 - 1)$
can be thought of as a measure of the specific kinetic energy,
we interpret this part of the increased
relativistic gravity as arising from
the increasing mass energy associated with the increasing
four velocity the binary pair as they approach.

Although, we expect that post-Newtonian relativity
is a poor approximation in the strong fields near the
neutron stars, it is nevertheless instructive to look
for this $(W^2 - 1)$ effect in
post-Newtonian relativity.  One motivation for
such an analysis is that
the post-Newtonian expansion is an
independent approximation to the metric.
Hence, it can help to
dispel concern as to whether the stellar collapse
is somehow an artifact of our metric choice.
This  exercise can also provide some intuitive
insight as to the origin of this phenomenon.
In this appendix, therefore,
we outline how a simple post Newtonian expansion
can exhibit an enhancement of the effective gravitational potential
which involves terms scaling  as  $(W^2 - 1)$.

  In the post-Newtonian approximation (cf. Weinberg~1972)
one presumes that it is possible to write the metric
as the sum of the Minkowski tensor plus corrections
given by an expansion in powers of $v^2 \sim (GM/r)$ in a conformally
flat metric.  For example,
the time-time component of the metric is written,
\begin{equation}
g_{t t} = -1 +  g_{t t}(2) +  g_{t t}(4)\cdots~~,
\end{equation}
where the numbers in parentheses $(n)$ denote terms of order of $v^n$.
From this metric, the corresponding components
of the affine connection and Ricci tensor can be similarly
expanded.  One also decomposes the stress energy tensor into
the Newtonian rest mass density plus corrections.
\begin{equation}
T^{t t} =  T^{t t}(0)  + T^{t t}(2) + T^{t t}(4) \cdots
\end{equation}
\begin{equation}
T^{i j} =  T^{i j}(2)   + T^{i j}(4) \cdots~~,
\end{equation}
where the numbers in parentheses now denote terms of order
$(M/r^3)v^n$ and
\begin{equation}
T^{t t}(0) \equiv  \rho(1 + \epsilon)~~.
\end{equation}
 The remaining $T^{\mu \nu}(n)$ terms then derive
from subtracting $T^{t t}(0)$ from
the perfect fluid tensor and retaining terms at
the appropriate order.

Let us consider  only the $g_{t t}$ component of the metric as an indicator
of the strength of the gravitational field.
For example, the first post-Newtonian  correction to the metric,
$g_{t t}(2)$ term, is just
\begin{equation}
\nabla^2g_{t t}(2) = -8 \pi G  T^{t t}(0)
\end{equation}
Thus, we have
\begin{equation}
g_{t t}(2) = -2 \Phi
\end{equation}
where $\Phi$ is just the Newtonian gravitational potential.  Velocity
dependent terms enter
at the next order.  Following Weinberg (1972) we have
\begin{equation}
g_{t t}(4) = -2 \Phi^2 - 2 \Psi~~~,
\end{equation}
where  $\Psi$ is a second gravitational potential given by
\begin{equation}
\nabla^2 \Psi = {\partial^2 \Phi \over \partial t^2} + 4 \pi G
[T^{t t}(2) +  T^{i i}(2)]~~~.
\label{psieq}
\end{equation}
To obtain insight into $\Psi$ we first write the appropriate
terms from the perfect fluid energy
momentum tensor,
\begin{eqnarray}
T^{t t}(2) &=&   \rho(1 + \epsilon)\biggl[(W^2 - 1) - 2 \Phi \biggr]~~,
\end{eqnarray}
where the  approximate identification of $(W^2 - 1)$ with $v^2$
is valid at order $v^2$.
The spatial part is just
\begin{equation}
T^{i i}(2) =  ( W^2 -1)\rho(1 + \epsilon)  + 3 P~~.
\end{equation}
Thus, the source for the Laplacian of
$\Psi$ includes terms
which scale as $(W^2 - 1)$ times the mass-energy density plus
a smaller contribution from the second order time derivative 
of the Newtonian potential.
\begin{equation}
\nabla^2 \Psi = {\partial^2 \Phi \over \partial t^2} +  4 \pi G \biggl[
2 \rho(1 + \epsilon) (W^2 - 1 - \Phi)  +  3 P \biggr]
\end{equation}
This  suggests how the effective 
gravitational potential might be deeper for binary stars (where 
$(W^2 - 1) > 0$) than the static potential of two isolated stars.
The total compression effect, however, derives from the covariant derivative
of the stress energy tensor as illustrated in Eq. \ref{hydrostat}.
One must consider the effective hydrostatic equilibrium of each star
which may involve terms of higher order than first post Newtonian.

\end{document}